\documentclass[aps,prl,twocolumn,showpacs,superscriptaddress,floatfix,reqno]{revtex4-2} 
\usepackage{graphicx}
\usepackage{dcolumn}
\usepackage{bm}
\usepackage{hyperref}
\usepackage[utf8]{inputenc}
\usepackage[english]{babel}
\usepackage{units}
\usepackage{mathtools}
\hypersetup{colorlinks=true,citecolor={blue},linkcolor={blue},urlcolor={blue}}
\usepackage{float}
\usepackage{amsmath}
\usepackage{empheq}
\usepackage{mathrsfs}
\usepackage{amssymb}
\usepackage{soul}
\usepackage{textcomp}
\usepackage{placeins}
\usepackage{xcolor}
\usepackage[T1]{fontenc}
\usepackage{tipa}
\setstcolor{red}

\usepackage[]{mdframed}
\begin{document}

\title{Hong-Ou-Mandel interferometry with trapped polariton condensates}

\author{S. Baryshev}
\email{stepan.baryshev@skoltech.ru}
\address{Skolkovo Institute of Science and Technology, Moscow, Territory of innovation center “Skolkovo”, Bolshoy Boulevard 30, bld. 1, 121205, Russia.}

\author{I. Smirnov}
\address{Skolkovo Institute of Science and Technology, Moscow, Territory of innovation center “Skolkovo”, Bolshoy Boulevard 30, bld. 1, 121205, Russia.}

\author{I. Gnusov}
\address{Skolkovo Institute of Science and Technology, Moscow, Territory of innovation center “Skolkovo”, Bolshoy Boulevard 30, bld. 1, 121205, Russia.}

\author{T. Cookson}
\address{Skolkovo Institute of Science and Technology, Moscow, Territory of innovation center “Skolkovo”, Bolshoy Boulevard 30, bld. 1, 121205, Russia.}

\author{A. Zasedatelev}
\address{Skolkovo Institute of Science and Technology, Moscow, Territory of innovation center “Skolkovo”, Bolshoy Boulevard 30, bld. 1, 121205, Russia.}

\author{S. Kilin}
\address{National Academy of Sciences of Belarus, Minskaja voblasć, Minsk, prasp. Niezaliežnasci 66, 220072, Belarus}

\author{P. G. Lagoudakis}
\address{Skolkovo Institute of Science and Technology, Moscow, Territory of innovation center “Skolkovo”, Bolshoy Boulevard 30, bld. 1, 121205, Russia.}

\begin{abstract}
We investigate the indistinguishability of polaritons in optically trapped Bose Einstein condensates by implementing Hong-Ou-Mandel (HOM) interferometry and test the limitations of two-polariton interference in the coherent, limit-cycle and thermal statistical regimes. We observe that the HOM dynamics of a circularly polarized condensate follows the condensate coherence time with the characteristic HOM-dip approaching the classical limit. Linearly polarized condensates exhibit a combined effect of polariton bunching and two-polariton interference. Under elliptically polarized excitation, the temporal evolution of the spinor condensate results in the revival of the HOM-dip at the spinor Larmor precession frequency. 


\end{abstract}

\maketitle
 

%
%
The Hong-Ou-Mandel two-particle interference (HOM TPI) effect marked the dawn of a new era of quantum progress with its fundamental manifestation of the bosonic properties of light~\cite{PhysRevLett.59.2044}. 
The observation of quantum interference with massive particles has enabled the possibility to test the theories of the quantum-to-classical transitions ~\cite{RevModPhys.76.1267}, and Bell's inequality~\cite{PhysRevA.91.052114}. 
Quantum interference was observed in atomic Bose-Einstein condensates (BEC) ~\cite{doi:10.1126/science.1250057, Lopes2015}. Beyond the physics of atomic BEC, condensates of exciton-polaritons (here on polaritons) offer strong interactions permitting the observation of squeezing~\cite{Boulier2014}, quantum correlations~\cite{Delteil2019, Matutano2019} and long range coherent coupling~\cite{PhysRevLett.124.207402}. The condensates of polaritons are extensively studied as a potential platform for the study of quantum fluids of light~\cite{Amo2009,superfl} and quantized vortices~\cite{doi:10.1126/sciadv.add1299} as well as for a variety of device applications~\cite{polaritonicdevices}. On chip switchable polaritonics~\cite{Zasedatelev2021}, simulators~\cite{AMO2016934,Berloff2017} and logic gates~\cite{PhysRevB.85.235102, Zasedatelev2019} are at pinnacle of community research interest.

\par The recent observation~\cite{PhysRevLett.128.087402} on different photon statistical regimes encouraged us to investigate particle-interference in an optically trapped polariton condensate. We follow the HOM effect through the crossover from fully coherent to thermal condensate statistics. We explore how different photon statistics of the condensate spinor projection affect the landscape of correlations in the HOM effect. We observe the HOM visibility revivals and show its dependence on the spinor Larmor precession frequency. The developed theoretical model describes the experimental observation and predicts the behavior of correlation functions in the different photon statistical regimes.

\par  In our experiments, polaritons are generated through strong light-matter coupling of excitons with photons confined in a microcavity. They can undergo non-equilibrium Bose–Einstein condensation~\cite{Kasprzak} with a spinor order parameter $\Psi = (\psi_+,\psi_-)^T$ corresponding to the right-hand $\sigma^+$ and left-hand $\sigma^-$ circular polarization of the emitted light.  Interaction of polaritons with an incoherent excitonic reservoir has been shown to limit the coherence time of the condensate and induce depolarization~\cite{askitopoulos2019giant,PhysRevB.102.125419}. Methods of optical excitation that facilitate the separation of the excitation area from the region of condensate formation, while simultaneously providing a trapping potential, have proven to be effective in increasing the coherence time of polariton condensates. Furthermore, the tunable parameters of optical excitation such as pump power and polarization, along with the optical trap size and geometry are few of many degrees of freedom for controlling the condensate internal spinor dynamics. When excited with linearly polarized light, the local birefringence in the microcavity can cause linear polarization energy splitting within the condensate, analogous to a local in-plane effective magnetic field~\cite{KLOPOTOWSKI_SSC2006}. In such a case, polariton-polariton interactions align the condensate polarization to the effective field~\cite{PhysRevB.80.195309,PhysRevB.75.045326} and as a consequence photon statistics for the suppressed polarization direction becomes highly bunched. Additionally, by imprinting a spin population imbalance in both the condensate and the exciton reservoir it is possible to induce an effective out-of-plane magnetic field which leads to the self-induced Larmor precession of the condesate spinor~\cite{Solnyshkov2007, Ryzhov_PhysRevRes2020, PhysRevLett.128.087402}.

\begin{figure}[t!]
\includegraphics [width=1\columnwidth]{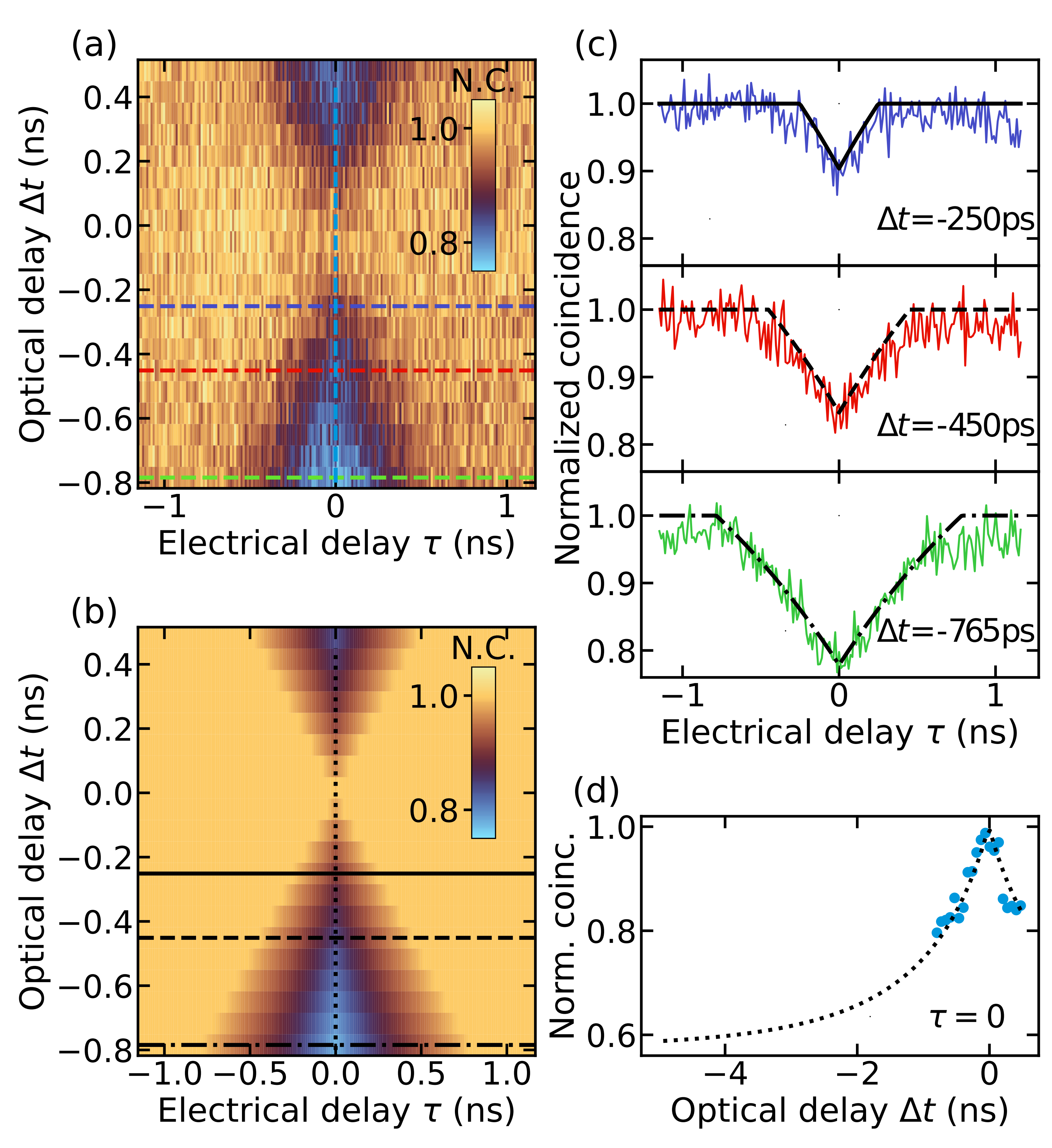}
\centering
\caption{ 
 (a) Color map of measured trapped condensate normalized intensity correlation functions (N.C.) exhibiting HOM dip along the optical delay $\Delta t$ at $P=1.13P_{th}$, excited with circularly polarized light. Dashed colored horizontal lines indicate the cross-sections for the data presented in panel (c), dashed vertical line indicate the cross-section for the data presented in panel (d).  
 (b) Corresponding simulated with Eq.~\ref{eq:7} condensate HOM dip color map. Black stylized horizontal lines indicate the cross-sections for the data presented in panel (c), dashed vertical line indicate the cross-section for the data presented in panel (d). 
 (c) Measured HOM dip for optical delays of -250ps (purple), -450ps (red), -765 ps (green) and corresponding simulation curves (black stylized lines) with respect to electrical delay $\tau$.
 (d) Measured depth of HOM dip at $\tau=0$ (blue circles) and corresponding theoretical prediction (dash dotted line) with the respect to optical delay $\Delta t$.}
\label{fig.1}
\end{figure}

\par In our work the optically trapped polariton condensate is created in a GaAs/AlAs$_{0.98}$P$_{0.02}$ 2$\lambda$-microcavity with embedded InGaAs quantum wells~\cite{doi:10.1063/1.4901814}, with the exciton-cavity detuning of $-3$~meV~\cite{PhysRevB.102.125419}. The microcavity is held in a cryostat at 4K and is excited non-resonantly at the first Bragg minimum of the reflectivity stop-band, $\lambda_\text{ext}=783$~nm. The continuous wave excitation is acousto-optically modulated at 1~kHz frequency and 10\% duty cycle and spatially modulated into an annular optical trap of $d=9.5\mu m$ in diameter. The polariton condensate forms in the center of the optical trap, minimizing the spatial overlap with the pump region. For all the trap sizes and excitation powers presented in this letter the condensation happens into a single energy level which corresponds to the optical trap ground state. The condensate emission is directed into the Mach-Zehnder type interferometer with an arm extension, which provides the optical time delay $\Delta t$. The interferometer outputs are coupled into single photon counters connected to the correlation electronics giving access to electrical time delay between photon detection $\tau$. Both inputs have the same polarization. See Supplemental Information (SI) Sec.1 for the experimental setup.

        \par We begin from the derivation of a general equation for the second-order correlation function $g^{(2)}_{HOM}$ which we would later obtain through the measurement. The result of the field interference $E_{D1,D2}(t)$ at the detectors can be expressed as:

        
        \begin{equation}\label{eq:1}
            E_{D1,D2}(t)=\frac{1}{2}(E(t)\pm E(t+\Delta t)).
        \end{equation}

        The photon statistics in the HOM interferometer correspond to the following second-order correlation function:
        
        \begin{equation}\label{eq:2}
        G^{(2)}_{HOM}(\tau,\Delta t)=\Big< E_{D1}^*(t)E_{D2}^*(t+\tau)E_{D2}(t+\tau) E_{D1}(t)\Big>,
        \end{equation}

        where $<>$ is time-averaging over millions of condensate realizations (system is ergodic). When expanded, Eq.~\ref{eq:2} contains 16 terms and can be simplified, when we average the phase dependent terms. The detailed derivation can be found in SI Sec.2. The equation becomes as follows:
        
        \begin{equation} \label{eq:3}
          G^{(2)}_{HOM}(\tau,\Delta t)=\frac{1}{16}\Big(G^{(2)}(\tau) +G^{(2)}(\tau-\Delta t)+ 
        \end{equation}
        $$
            +G^{(2)}(\tau+\Delta t)+G^{(2)}(\tau) \Big)-\frac{1}{8}G^{(2)}(0,\Delta t+\tau,  \tau,\Delta t). 
        $$

        All the correlation functions are normalized onto the level of long electrical delays $\tau$ where any statistical features are absent. The corresponding normalized version of Eq.~\ref{eq:3} reads:
        
        \begin{equation} \label{eq:ME}
            g^{(2)}_{HOM}(\tau,\Delta t)=\frac{ G^{(2)}_{HOM}(\tau,\Delta t)}{G^{(2)}_{HOM}(\infty,\Delta t)}
        \end{equation}
        
        
        The Eq.~\ref{eq:ME} describes the HOM effect in the general case of an interferometer without the phase stabilization. $G^{(2)}(\tau)$ can be measured directly by closing one of the arms of the interferometer, converting it into Hanbury-Brown and Twiss (HBT) interferometer. The $G^{(2)}(0, \Delta t+ \tau,\tau,\Delta t)$ term, however cannot be measured directly and requires the calculations of the developed theoretical model.




In Fig.~\ref{fig.1}(a) we show the normalized intensity correlation functions measured within the HOM interferometer versus optical $\Delta t$ and electrical $\tau$ delays. Here the polariton condensate was prepared at $1.13P_{th}$ with circularly polarized excitation, where $P_{th}$ is the condensation threshold. Polariton condensate under such conditions exhibit features similar to a single mode laser; narrow linewidth, high degree of polarization and Poissonian photon statistics with intensity autocorrelation function $g^{(2)}(0)=1$~\cite{Kasprzak,PhysRevLett.128.087402,PhysRevB.102.125419}. 
Accordingly, we describe the condensate PL classically:
        \begin{equation} \label{eq:5}
            E(t)=|E_0|e^{i\omega_0 t + i\varphi(t)},
        \end{equation}
        where $\omega_0$ is the PL central frequency and $\varphi(t)$ is a stochastic time-dependent phase that determines the linewidth. Here the first-order coherence function is:

         \begin{equation} \label{eq:6}
            g^{(1)}(\tau)=\frac{<E^*(t+\tau)E(t)>}{|E_0|^2}=\gamma e^{-|\tau|/\tau_0},
        \end{equation}
        where $\tau_0$ is the coherence time, and $\gamma\approx0.92$ is a factor designed to take into account the presence of incoherent radiation of uncondenced polaritons (see SI Sec.2A).
        

        As mentioned above the photon statistics of circularly polarized condensate PL slightly above the threshold is Poissonian, the first part of Eq.~\ref{eq:3} becomes equal to $|E_0|^4$, thus the shape of the HOM dip is determined by the last term. The final expression for the HOM interference in this case reads as:

         \begin{equation} \label{eq:7}
            g^{(2)}_{HOM}(\tau,\Delta t)=\frac{1-\frac{1}{2}|g^{(1)}(t_{min})|^2}{1-\frac{1}{2}|g^{(1)}(\Delta t)|^2},
        \end{equation} 
        where $t_{min}$ is the smallest between $|\tau|$ and  $|\Delta t|$.


        Figure~\ref{fig.1}(b) shows a good agreement of developed theoretical model to the experimentally obtained HOM effect color map. Color and style coded cross-section match in Fig.~\ref{fig.1}(c) exemplifying the temporal dynamics of the measured HOM dip versus electrical $\tau$ and as shown in Fig.~\ref{fig.1}(d), optical $\Delta t$ delays. The polarization instability between the inputs would be reflected as a decreased visibility of the effect (see SI Sec.3). We highlight that when the two inputs to the beam splitter are coherent ($\Delta t=0$) the Eq.~\ref{eq:7} predicts no anti-correlations at $\tau=0$ same as it is observed in the experiment. The evolution of HOM dip towards $|\Delta t| \gg \tau_0$  reproduces the known result for classical sources~\cite{PhysRevA.28.929,Graefe_2008} $g^{(2)}_{HOM}(\tau)=1-\frac{1}{2}|g^{(1)}(\tau)|^2$ with a HOM dip equal to $\frac{1}{2}$. For the range of optical delays in our experiment, Eq.~\ref{eq:7} produces a curve with a dip at $\tau=0$ shaped as $-|g^{(1)}(\tau)|^2$ truncated by the optical delay $\Delta t$. The coherence time of optically trapped condensates is sensitive to the optical trap size geometry and respectively to the overlap of condensate with the exciton reservoir~\cite{askitopoulos2019giant}, which consequently affects the HOM effect (see SI Sec.4). Our measurements, together with theoretical confirmation matches classically expected values of HOM dip. This fortifies the claims of high spin stability~\cite{PhysRevB.102.125419} and underline the absence on dynamical instabilities~\cite{PhysRevB.90.205304} within the polariton condensate, which otherwise would cause spatial mode and polarization mismatch, diminishing the HOM effect.

\par Previous studies show, that the condensate spinor enters a dynamical regime of self-induced Larmor precession when excited with elliptically polarized optical trap~\cite{Solnyshkov2007,PhysRevLett.129.155301,PhysRevLett.89.077402}. The spin imbalance in the exciton reservoir is transferred as a spin imbalance within the condensate, resulting in an energy splitting and the  emergence of an effective out-of-plane magnetic field $\boldsymbol{\Omega}_\perp = \Omega_z \hat{\mathbf{z}}$~\cite{PhysRevLett.89.077402,Shelykh_PRB2004, Krizhanovskii_PRB2006, Solnyshkov2007, Ryzhov_PhysRevRes2020}. The field can be expressed as $\Omega_z\propto \alpha S_3 + g (X_+ - X_-)$, where $\alpha$ denotes the polariton-polariton interaction strength, $g$ the polariton-exciton interaction strength, and $X_\pm$ are the exciton reservoir spin populations. The direction and strength of the field can be controlled by tuning the polarization ellipticity of the pump laser~\cite{PhysRevB.99.165311, PhysRevB.102.125419}. The spinor precession projected on the equatorial plane of the Poincare sphere can be observed as periodic oscillations of intensity of the linear polarization components. This effect is evidenced in a periodic modulation of the intensity correlation function at the Larmor precession frequency for any of measured linear polarization projections. Frequency of such precessions can be tuned up to several GHz with spin dephasing times of up to 10ns~\cite{PhysRevLett.128.087402}, and at least hundreds of ns with an external drive~\cite{Gnusov:24}. 

        In order to model the $g^{(2)}_{HOM}(\tau,\Delta t)$ in such case we take into account the polariton condensate spin imbalance ($\psi_+\neq \psi_{-}$) induced by elliptically polarized excitation. We introduce electric field in the orthogonal polarization projections $E_{H,V}(t)$ as follows:
        
        \begin{equation} \label{eq:8}
             \begin{pmatrix} 
            E_H(t) \\ E_V(t)
            \end{pmatrix} =|E_0|e^{ i\varphi(t)+i\varphi_0(t)}\Big[
            \begin{pmatrix} 
            1 \\ i
            \end{pmatrix} 
           e^{i\omega_+ t} + r
            \begin{pmatrix} 
            1 \\ -i
            \end{pmatrix} 
            e^{i\omega_- t}\Big],
        \end{equation}
        where $\omega_+$ and $\omega_-$ are $\psi_+$ and $\psi_-$ natural frequencies, and the splitting $\Delta \omega=\omega_+-\omega_-$ corresponds to the frequency of Larmor oscillations that are observable in polarization filtered intensity correlation functions (see SI Sec.2). The $r$ factor reflects the mode imbalance $(0\leq r \leq 1)$. The width of $\omega_{+,-}$ is referred to as a phase noise $\varphi(t)$ which causes the decoherence of the condensate PL, while the width of $\Delta\omega$ is explicitly responsible for the dephasing of the Larmor precessions. The phase term $\varphi_0(t)$ is required to describe the additional modulation of the correlation functions which emerge at higher excitation densities $>3P_{th}$, i.e. in the range where Larmor precessions are prominent. 

        %
        
        In the experiment the measured polarization component is set to be horizontal, accordingly we consider only $E_{H}(t)$ from Eq.~\ref{eq:8}. The Eq.~\ref{eq:ME} for elliptically excited case reads:

        \begin{equation}
        \label{eq:9}
        \begin{split} 
            G^{(2)}_{HOM}(\tau,\Delta t)=\frac{1}{4}|E_0|^4 \Big[(1+r^2)^2\Big(1-\frac{1}{2}|g^{(1)}(t_{min})|^2\Big)+\\
            +2r^2 \sin^2 (\frac{\Delta \omega \Delta t}{2}) |g^{(1)}(t_{min})|^2+\\
            +2r^2 \cos(\Delta \omega \tau) \Big(\cos^2 (\frac{\Delta \omega\Delta t}{2}) -\frac{1}{2}|g^{(1)}(t_{min})|^2\Big)\Big]. 
        \end{split}        
      \end{equation}

        In order to account for the mentioned additional low frequency modulations the Eq.~\ref{eq:9} has to be augmented with additional term $K(\tau,\Delta t)$ (see SI Sec.2B). Although, it goes out of the scope of this work, such modulations could be caused by similar spectral features as in recent works on time crystal in a polariton condensate~\cite{doi:10.1126/science.adn7087}.

\begin{figure}[t!]
\includegraphics [width=1\columnwidth]{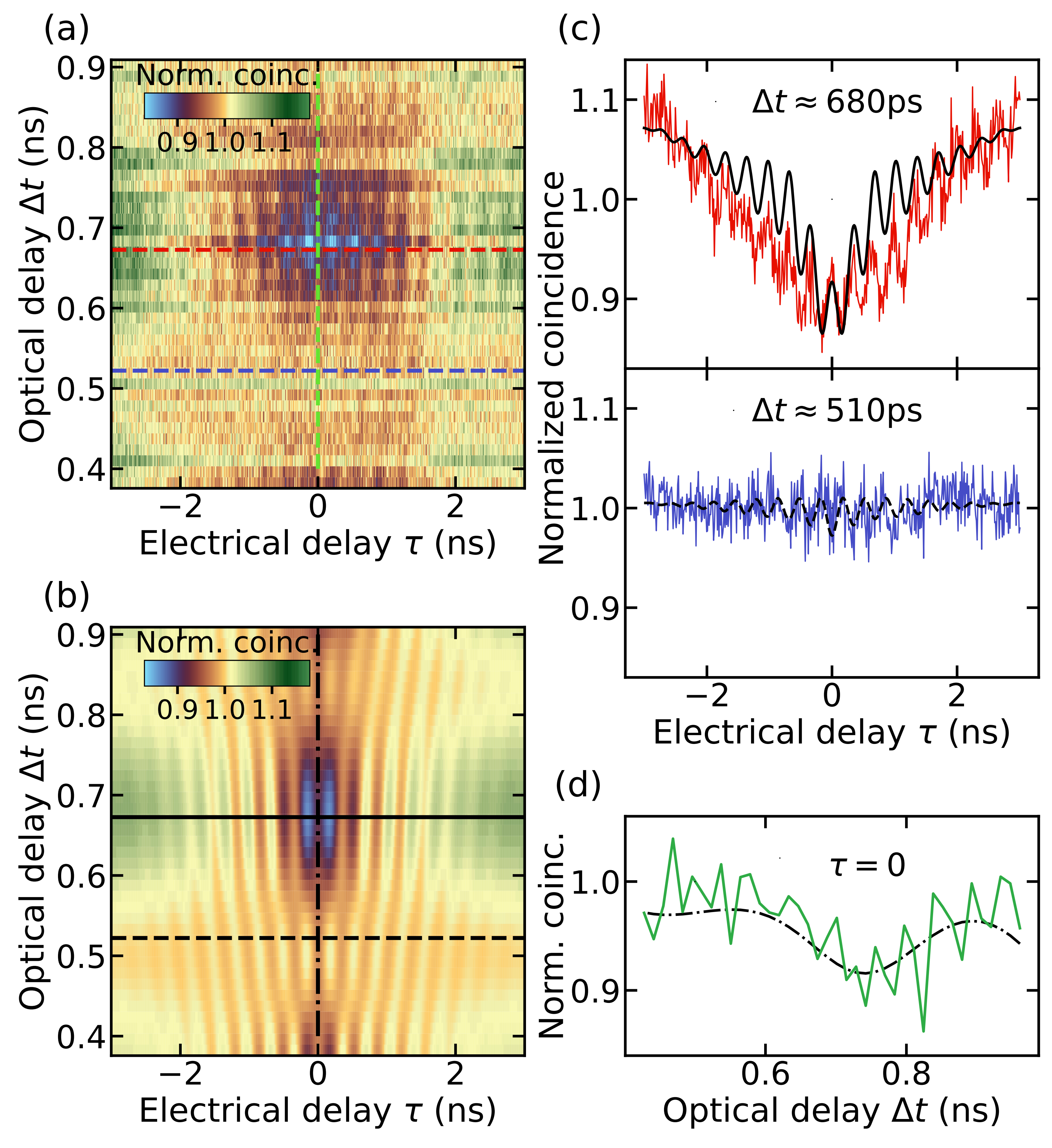}
\centering
\caption{(a) Measured condensate HOM dip color map for different positions of optical delay $\Delta t$ with elliptically polarized excitation. Dashed colored horizontal lines indicate the cross-sections for the data presented in panel (c), dashed vertical line indicate the cross-section for the data presented in panel (d).  
(b) Simulated according to Eq.~\ref{eq:9} condensate HOM dip color map. Black stylized horizontal lines indicate the cross-sections for the data presented in panel (c), dashed vertical line indicate the cross-section for the data presented in panel (d).  
(c) Profiles extracted along the horizontal cross-sections in panel (a) and  (b), showcase the visible HOM dip together with correlation function oscillations due to Larmor precessions.  
(d) Profiles extracted along the vertical cross-sections in panel (a) and  (b) demonstrate the revival of the HOM dip visibility, due to precession phase matching.}
\label{fig.2}
\end{figure}

\par By managing the optical delay between the arms of the HOM interferometer it is possible to bring the periodic oscillations of photon number in and out of phase with respect to each other. The experimentally acquired color map and theoretically modeled confirmation in Fig.~\ref{fig.2}(a) and (b) provides the evidence of HOM effect revival for optical delays at which intensity oscillations are in phase $\Delta t \approx 680 ps$, and the absence of characteristic dip at $\Delta t  \approx  510 ps$. When the optical delay between the interferometer arms is equivalent to the integer number of the spinor Larmor precession periods, then the joint effect manifests itself as combined features of precession induced correlation function oscillation with a sharp peak overlapped with a broad HOM dip at $\tau=0$, $g^{(2)}_{HOM}(0,\Delta t)$, as shown in Fig.~\ref{fig.2}(c). The precession period for this data set is $T\approx340ps$, thereby revival of the HOM effect along optical delay  happens with the same periodicity, with half period of $T/2\approx170ps$ as shown in Fig.~\ref{fig.2}(d). 

The relief of the typical revival map, such as shown in Fig.~\ref{fig.2}(a), is sensitive to plethora of parameters. The Larmor precession frequency is directly connected to the HOM revival periodicity, and any change in the optical path will lead to the temporal shift of the set map. While the results presented here are taken with two co-aligned polarizers before the fiber beam splitter, it is possible to measure the similar HOM revival map by incorporating arbitrary alignment of polarizers, given that input polarization is matched after the polarizers to achieve maximum photon indistinguishably.

\par Under linearly polarized excitation the polariton condensate in an optical trap can to act as a source of highly bunched photons~\cite{PhysRevLett.128.087402}. The condensate can experience polarization mode competition and the collapse of system U(1) symmetry depending on cavity internal birefringence~\cite{PhysRevB.80.195309} and confining potential ellipticity~\cite{PhysRevApplied.16.034014,PhysRevB.99.115303}, which leads to a linear polarization energy splitting analogous to an in-plane effective magnetic field $\boldsymbol{\Omega}_\parallel$. This results in 'pinning' of one of the polarization modes to some preferred direction in the cavity plane, while the gain for the orthogonal mode gets strongly suppressed. Because of this the intensity correlation function for the suppressed polarization mode of polariton condensate reach values $g^{(2)}(0)>1$ above condensation threshold. The amplitude of the correlation function, as well as its characteristic decay times, can be manipulated through the parameters of in-plane magnetic field $\boldsymbol{\Omega}_\parallel$, and by providing enough gain to the system saturating all the polarization modes. 

\par Polarization selective HOM measurements unavoidably lead to the observation of the combined effect of high $g^{(2)}(0)$ values for the suppressed polarization mode and the dip due to HOM effect. In the following measurements, we fix the optical delay between the paths of the HOM interferometer at $\Delta t=650ps$. In such a case, since the photons are bunched ($g^{(2)}(0)>1$) the description of obtained results can be done assuming thermal statistics of condensate PL. It is common knowledge that the higher-order correlators can be expressed with a single correlator of the first order~\cite{MICHALOWICZ20111233,klauder2006fundamentals}. By this virtue, we can omit the calculations of multi-time correlators, the mode-spectral composition of radiation, and encapsulate all decoherence mechanisms into $g^{(1)}(\tau)$ analogously to Eq.~\ref{eq:7}.

\begin{figure}[t!]
\includegraphics [width=1\columnwidth]{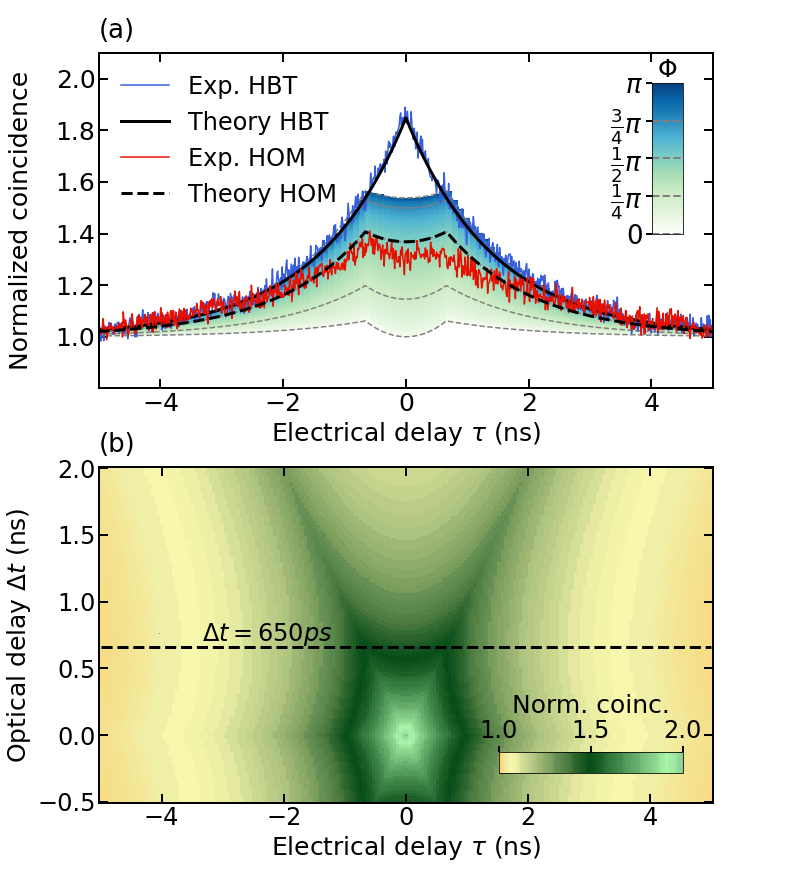}
\centering
\caption{(a) Measured with HBT interferometer photon bunching for horizontally polarized condensate PL (blue), and corresponding correlation function measured with HOM interferometer (red). The theoretical predictions with Eq.~\ref{eq:10} for HBT and~\ref{eq:11} for HOM interferometers are shown as the black solid and dashed lines, respectively. The color gradient depicts the predicted with Eq.~\ref{eq:12} normalized coincidences of HOM effect with interferometer stabilized at different phase shifts ($\Phi$). Grey dashed lines are to guide the eye and represent correlation functions at phase shifts $\Phi$ values denoted on the color bar.
(b) Simulated HOM effect color map for different optical delays. The black dotted line indicates the cross section corresponding to the measured (red) and predicted (dashed) HOM curves on the panel (a).}

\label{fig.3}
\end{figure}

        With this assumption, the following formula can be obtained for HBT measurements:
        
        \begin{equation} \label{eq:10}
             g^{(2)}_{HBT}(\tau)=  1+\gamma^2 e^{-2|\tau|/\tau_0}.
        \end{equation}

        Finally, Eq.~\ref{eq:ME} in the case of linearly polarized excitation reads as follows:
        
        \begin{equation} \label{eq:11}
            g^{(2)}_{HOM}(\tau,\Delta t)= 1+\frac{1}{4}\frac{|g^{(1)}(\Delta t+\tau)|^2 + |g^{(1)}(\Delta t-\tau)|^2}{1-\frac{1}{2} |g^{(1)}(\Delta t)|^2}.
        \end{equation}

        
        Additionally, we calculate the shape of the correlation function, shown as a gradient, when the phase shift $\Phi$ between interferometer arms is fixed, i.e. $\Delta t$ does not fluctuate. The following expression holds:
        
        \begin{equation} \label{eq:12}
        \begin{split}
            & g^{(2)}_{HOM}(\tau,\Delta t, \Phi)= 1+\\
        +\frac{1}{4} & \frac{|g^{(1)}(\Delta t+\tau)|^2 + |g^{(1)}(\Delta t-\tau)|^2}{1-\frac{1}{2}(1+\cos(2\Phi)) |g^{(1)}(\Delta t)|^2}-\\        
        -\frac{1}{2} & \frac{ \cos (2\Phi)
        g^{(1)}(\Delta t-\tau) g^{(1)}(\Delta t+ \tau)}{1-\frac{1}{2}(1+\cos(2\Phi)) |g^{(1)}(\Delta t)|^2}.  
        \end{split}
\end{equation}

In Fig.~\ref{fig.3}(a) we show the photon bunching of the suppressed polarization mode (horizontal) with its intensity correlation function $g^{(2)}(0)\approx1.88$, at $P=1.4P_{th}$. The measured HOM correlation function $G^{(2)}_{HOM\parallel}(\tau)$, is shown in the same plot in blue. At longer delays ($\tau>2ns$), the correlation function follows the characteristic decay of photon bunching. However, near $\tau=0$ the photon statistics is severely affected by the HOM TPI effect. The developed theoretical model convincingly matches the experimental data and shows the possibility to further control the HOM effect by stabilizing the interferometer and fixing the phase shift $\Phi$. The Fig.~\ref{fig.3}(b) demonstrates how the shape of the HOM effect would evolve with increasing optical delay. Due to the high bunching at $\tau=0$, the HOM dip is twice as pronounced in absolute values if compared to the case of PL with Poissonian statistics.

\par In conclusion, we studied the HOM effect of optically trapped polariton condensate spinor projection in three distinct photon statistical regimes conditioned by the polarization ellipticity of the non-resonant excitation. We developed a theoretical model which comprehensively describes the observed statistical features. We started with the simplest case of a circularly polarized condensate and showed how condensate coherence time correspond to HOM dynamics. We estimate HOM correlations for such condensate with the theoretical model for classical light sources underlining the high degree of condensate spin co-circularity and the absence of dynamical instabilities for condensate spatio-temporal distribution. By tuning the ellipticity of the excitation laser we were able to control the frequency of condensate spinor precessions which formate the landscape of HOM effect causing revivals of HOM dip at the Larmor frequencies. Additionally we have observed the low frequency modulation of the HOM landscape, which could be caused by the spectral features reminiscent of time crystals. Finally, we showed that the absolute HOM dip is twice as pronounced when the condensate is excited with linearly polarized light and photon statistics is thermal. The phase sensitivity was predicted to greatly affect the HOM dip, with amplitude averaging into observed value in a non-stabilized interferometer. 

This work was supported by Russian Science Foundation (RSF) grant no. 24-72-10118, https://rscf.ru/en/project/24-72-10118/.

\newpage

\setcounter{equation}{0}
\setcounter{figure}{0}
\setcounter{section}{0}
\renewcommand{\theequation}{S\arabic{equation}}
\renewcommand{\thefigure}{S\arabic{figure}}
\renewcommand{\thesection}{S\arabic{section}}
\onecolumngrid

\vspace{1cm}
\begin{center}
\Large \textbf{Supplemental Information: Hong-Ou-Mandel interferometry with trapped polariton
condensates}
\end{center}

\section{Experimental setup}

Microcavity sample at 4K is excited non-resonantly ($\lambda_{ext}=783nm$) into cavity Bragg minimum with annular beam profile formed with spatial light modulator (SLM). The excitation polarization ellipticity is controlled with a quarter waveplate (QWP). Optically trapped condensate PL is directed into a HOM interferometer where BS1 and BS2 are 50:50 free-space, and single mode fiber beamsplitters respectively. Polarizers (P1,2) are aligned to match the polarizations on BS2, while additional polarization control with half waveplate (HWP) and QWP is used to maintain the polarization matching in the fiber. Single photon detectors (SPAD1,2) register the photon arrival. Movable mirror provides optical delay $\Delta t$ and correlator electronics electrical delay $\tau$.

\begin{figure}[h!]
\includegraphics [width=0.6\columnwidth]{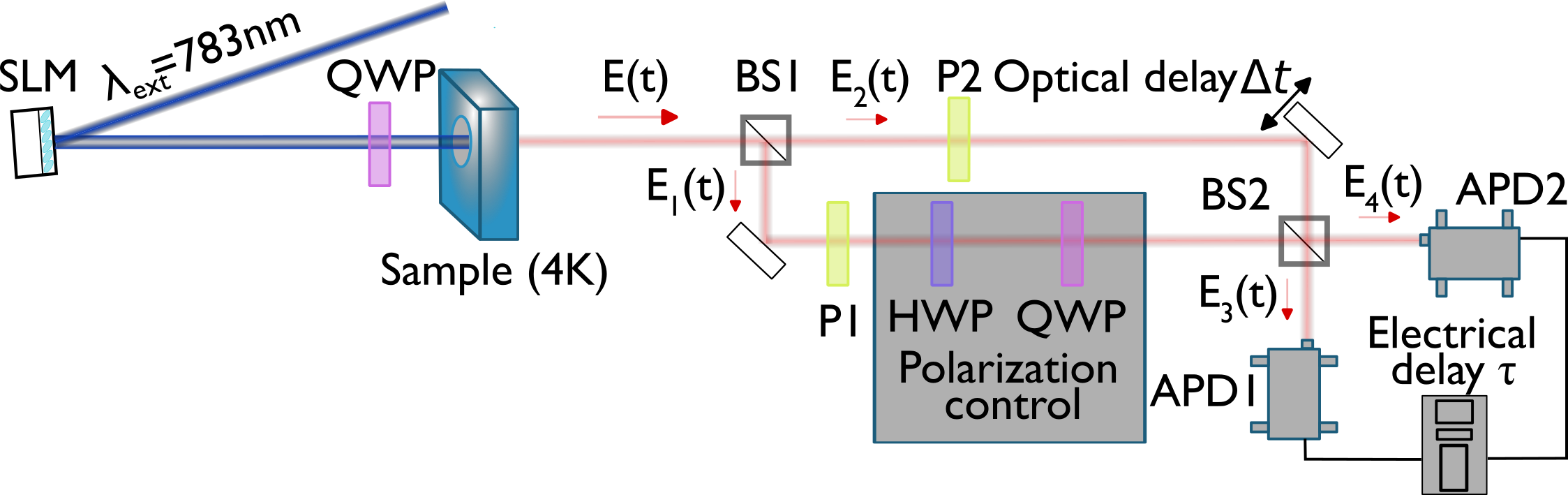}
\centering
\caption{Schematic of the experimental setup.}
\label{fig.1}
\end{figure}

\section{Theoretical model}

To describe the HOM effect, it is possible to derive the master equation that is valid for any properties of emission. Namely, let the sample emit a field $E(t)$.
At the input to the interferometer, the polarizer $P$ filters out only the horizontal emission mode. In further calculations, the field will be understood as a stochastic $c-$number function, however the expression obtained further is retained for description using operators. 
According to Fig 1(a), next the field $E(t)$ falls on BS and is split into two components 
$E_{1,2}(t)=\frac{1}{\sqrt{2}}E(t),$ which propagate along the arms of the interferometer of different lengths, acquiring a controlled phase difference. Thus, at the entrance to the second BS, without loss of generality, the fields can be written as $E_{1}(t)=\frac{1}{\sqrt{2}}E(t)$ and $E_{2}(t)=\frac{1}{\sqrt{2}}E(t+\Delta t)$, where $\Delta t$ denotes the optical time delay. Finally, at the output of the second BS, the fields are obtained:
\begin{equation}
    E_{3,4}(t)=\frac{1}{2}(E(t)\pm E(t+\Delta t))
\end{equation}

These fields are collected using fibers and recorded using two single-photon APD detectors connected to a TCSPC single-photon counter that records the time difference between the events, referred to as electrical time delay $\tau$.

In such a scheme, the measured statistics correspond to the following expression for the second-order correlation function:

\begin{equation}
    G^{(2)}_{HOM}(\tau,\Delta t)=\Big< E_3^\dagger(t)E_4^\dagger(t+\tau)E_4(t+\tau) E_3(t)\Big>
\end{equation}

Here $\dagger$ corresponds to Hermitian conjugation, and $<>$ corresponds to time averaging over numerous condensate realizations. Substituting (1) into (2), we obtain the expression:

\begin{equation}
    G^{(2)}_{HOM}(\tau,\Delta t)=\frac{1}{16}\Big<\big(E^\dagger(t)+E^\dagger(t+\Delta t)\big)\big(E^\dagger(t+\tau)-E^\dagger(t+\tau+\Delta t)\big)\times
\end{equation}
$$
\times\big(E(t+\tau)-E(t+\tau+\Delta t)\big)\big(E(t)+E(t+\Delta t)\big)\Big>
$$

Let's introduce the notation $G^{(2)}(t_1,t_2,t_3,t_4)=<E^\dagger(t_1)E^\dagger(t_2)E(t_3)E(t_4)>$. Note that due to the fact that the system is stationary, it is possible to put $t=0$ without loss of generality. Then equation (3) decomposes into 16 terms.

\begin{equation} \label{eq:4}
G^{(2)}_{HOM}(\tau,\Delta t)=\frac{1}{16}
\left[{
\begin{array}{lclr}
    +G^{(2)}(0,\tau,\tau,0)    & (1) & +G^{(2)}(\Delta t,\tau,\tau,0)   & (9)  \\
    +G^{(2)}(0,\tau,\tau,\Delta t)    & (2) &+G^{(2)}(\Delta t,\tau,\tau,\Delta t)   & (10)  \\
    -G^{(2)}(0,\tau,\tau+\Delta t,0)    & (3) & -G^{(2)}(\Delta t,\tau,\tau+\Delta t,0)   & (11)  \\
    -G^{(2)}(0,\tau,\tau+\Delta t,\Delta t)    & (4) & -G^{(2)}(\Delta t,\tau,\tau+\Delta t,\Delta t)   & (12)  \\
    -G^{(2)}(0,\tau+\Delta t,\tau,0)    & (5) & -G^{(2)}(\Delta t,\tau+\Delta t,\tau,0)   & (13)  \\
    -G^{(2)}(0,\tau+\Delta t,\tau,\Delta t)    & (6) & -G^{(2)}(\Delta t,\tau+\Delta t,\tau,\Delta t)   & (14)  \\
    +G^{(2)}(0,\tau+\Delta t,\tau+\Delta t,0)    & (7) & +G^{(2)}(\Delta t,\tau+\Delta t,\tau + \Delta t,0)   & (15)  \\
    +G^{(2)}(0,\tau+\Delta t,\tau+\Delta t,\Delta t)    & (8) & +G^{(2)}(\Delta t,\tau+\Delta t,\tau+\Delta t,\Delta t)   & (16)  \\ 
\end{array}}
\right]
\end{equation}

Note that since the electric field is proportional to the phase factor $E(t)\propto e^{-i\omega_0 t}$, where $\omega_0$ is the optical frequency, the correlation function is proportional to the following phase factor  $G^{(2)}(t_1,t_2,t_3,t_4) \propto e^{i\omega_0(t_1+t_2-t_3-t_4)}$. In the experiment, the interferometer is not stabilized, so the optical delay $\Delta t$ fluctuates in the sub-picosecond range, which leads to the cancelling of all terms proportional to the phase factor like $e^{-i\omega_0\Delta t}$.   Thus, the equation ~\ref{eq:4} is simplified to:

\begin{equation}
    G^{(2)}_{HOM}(\tau,\Delta t)=\frac{1}{16}
\left[{
\begin{array}{lclr}
    +G^{(2)}(0,\tau,\tau,0)    & (1) & +G^{(2)}(\Delta t,\tau,\tau,\Delta t)   & (10)  \\
    -G^{(2)}(0,\tau+\Delta t,\tau,\Delta t)    & (6) & -G^{(2)}(\Delta t,\tau,\tau+\Delta t,0)   & (11)  \\    
    +G^{(2)}(0,\tau+\Delta t,\tau+\Delta t,0)    & (7) & +G^{(2)}(\Delta t,\tau+\Delta t,\tau+\Delta t,\Delta t)   & (16)  \\ 
\end{array}}
\right].
\end{equation}

All terms except $(6)$ and $(11)$ are reduced to the directly observable autocorrelation function $G^{(2)}(\tau) = <E^\dagger(t)E^\dagger(t+\tau)E(t+\tau)E(t)>$, measured in the HBT scheme, while the terms $(6)$ and $(11)$ are complex conjugate. The system is symmetric to the rearrangement of detectors $(\tau\rightarrow - \tau)$ and to the rearrangement of paths in the interferometer $(\Delta t\rightarrow - \Delta t)$. Therefore, the term (6) turns out to be real since the following chain of equalities is valid:

\begin{equation}
    (G^{(2)}(0,\tau+\Delta t, \tau,\Delta t))^\dagger=G^{(2)}(\Delta t, \tau,\tau+\Delta t,0) = G^{(2)}(-\Delta t, \tau,\tau-\Delta t,0) = G^{(2)}(0, \tau+\Delta t,\tau,\Delta t).
\end{equation}

Thus the final expression can be written as:

\begin{equation} \label{eq:6}
  G^{(2)}_{HOM}(\tau,\Delta t)=\frac{1}{16}\Big(G^{(2)}(\tau)+G^{(2)}(\tau-\Delta t)+G^{(2)}(\tau+\Delta t)+G^{(2)}(\tau) \Big)-\frac{1}{8} G^{(2)}(0,\tau+\Delta t, \tau,\Delta t)   
\end{equation}

To describe the experimental data,  it is necessary to normalize  equation ~\ref{eq:6}:

\begin{equation} \label{eq:7}
    g^{(2)}_{HOM}(\tau,\Delta t)=\frac{G^{(2)}_{HOM}(\tau,\Delta t)}{G^{(2)}_{HOM}(\infty,\Delta t)}
\end{equation}

Note that in the experiment, each condensate realization lives on the order of $10\mu s$. Due to the fact that the optical delay $\Delta t$ fluctuates at frequencies characteristic of mechanical vibrations, it can be assumed that the phase difference $\Phi$ between the arms of the interferometer is fixed in each measurement. Thus, even for large electrical delays $\tau$, the intensity correlation across the diodes $<I_1(t+\tau,\Phi)I_2(t,\Phi) >\neq <I_1><I_2>$. In this regard, the normalization in~\ref{eq:7} is performed precisely for averaging product  $<I_1\cdot I_2>_\Phi$ over fluctuations of $\Phi$. In the following sections, we apply the equation ~\ref{eq:7} to describe three particular cases.


\subsection{Circularly polarized excitation}

In the case of circularly polarized excitation, condensate photoluminescence (PL) is co-circular with the excitation and in many ways resembles laser radiation. Well-defined intensity $(g^{(2)}(0) = 1)$ and narrow linewidth make possible the following quasi-classical description: 

\begin{equation} \label{eq:9}
    E(t)=|E_0|e^{-i\omega_0 t + i\varphi(t)},
\end{equation}

where $\omega_0$ is the angular frequency of the PL, $|E_0|$ is the fixed field amplitude, and $\varphi(t)$ is the stochastic time-dependent phase noise with independent increments, similar to the phenomenological  description of laser emission ~\cite{klauder2006fundamentals}. The properties of radiation coherence in this case are dictated by the following relation:

\begin{equation}
    g^{(1)}(\tau)=<e^{i\varphi(t+\tau)-i\varphi(t)}>=e^{-|\tau|/\tau_0}
\end{equation}

where $\tau_0$ is the coherence time of the condensate radiation. The decreased values of $g ^{(1)}(0)<1$ are observed in the experiment are associated with the presence of an incoherent component in the condensate PL. We take into account its presence by introducing a scale factor $\gamma$:

\begin{equation}
    g^{(1)}(\tau)=\gamma e^{-|\tau|/\tau_0}.
\end{equation}

In correspondence to the equation ~\ref{eq:9}, for the case of Poissonian statistics, we obtain:

\begin{equation}
    G^{(2)}(\tau)=\Big<E^\dagger(t)E^\dagger(t+\tau)E(t+\tau)E(t)\Big>=
\end{equation}
$$
=|E_0|^4\Big<e^{i\varphi(t)+i\varphi(t+\tau)-i\varphi(t+\tau)-i\varphi(t)}\Big>=|E_0|^4
$$

Thus, in the equation ~\ref{eq:6} the first 4 terms turn into a constant. Calculation of the last term gives:


\begin{equation}
    G^{(2)}(0,\tau+\Delta t,\tau, \Delta t) =\Big<E^*(t)E^*(t+\tau+\Delta t)E(t+\tau)E(t+\Delta t)\Big>=
\end{equation}
$$
=|E_0|^4\Big<e^{i\varphi(t)+i\varphi(t+\tau+\Delta t)-i\varphi(t+\tau)-i\varphi(t+\Delta t)}\Big>
$$

In the case of $|\Delta t|>|\tau|$, the increments $\varphi(t+\tau)-\varphi(t)$ and $\varphi(t+\Delta t+\tau)-\varphi(t+\Delta t)$ are independent and following transformation is legit:

$$
\Big<e^{i\varphi(t)+i\varphi(t+\tau+\Delta t)-i\varphi(t+\tau)-i\varphi(t+\Delta t)}\Big>=
\Big<e^{i\varphi(t)-i\varphi(t+\tau)} \Big> 
\Big< e^{i\varphi(t+\tau+\Delta t)-i\varphi(t+\Delta t)}\Big>=\left|{g^{(1)}(\tau)}\right|^2
$$

Conversely, if $|\Delta t|<|\tau|$ the increments $\varphi(t+\Delta t)-\varphi(t)$ and $\varphi(t+\Delta t +\tau)-\varphi(t+\tau)$ are independent and the following expression holds:

$$
\Big<e^{i\varphi(t)+i\varphi(t+\tau+\Delta t)-i\varphi(t+\tau)-i\varphi(t+\Delta t)}\Big>=
\Big<e^{i\varphi(t)-i\varphi(t+\Delta t)} \Big> 
\Big< e^{i\varphi(t+\tau+\Delta t)-i\varphi(t+\tau)}\Big>=\left|{g^{(1)}(\Delta t)}\right|^2
$$
Summarizing both point, the following expression holds true:
$$
G^{(2)}(0,\tau+\Delta t,\tau, \Delta t)= \begin{cases}
\begin{array}{ll}
    |E_0|^4 \left|g^{(1)}(\tau)\right|^2, & \text{for } |\Delta t|>|\tau|   \\
    |E_0|^4 \left|g^{(1)}(\Delta t)\right|^2, & \text{for } |\Delta t|<|\tau|   \\
\end{array} 
\end{cases}
$$

The result of calculations for ~\ref{eq:6} converges to:

\begin{equation}
    G^{(2)}_{HOM}(\tau,\Delta t)=\frac{1}{4}|E_0|^4\left({1-\frac{1}{2}\left|{g^{(1)}(t_{min})}\right|^2}\right)
\end{equation}

where $t_{min}=\min(|\Delta t|,|\tau|)$. The normalization factor is respectively equals to:

\begin{equation} \label{eq:15}
    \lim_{\tau\to\infty}G^{(2)}_{HOM}(\tau,\Delta t)=\frac{1}{4}|E_0|^4\left({1-\frac{1}{2}\left|{g^{(1)}(\Delta t)}\right|^2}\right)
\end{equation} 

Indeed, for large electrical delays $\tau$ the intensities on the diodes are equal to:
$$
I_{3,4}=\left<{|E_{3,4}|^2}\right>=\frac{1}{2}|E_0|^2\pm \frac{1}{2} \mathrm{Re} \left<{E^\dagger(t)E(t+\Delta t)}\right>=\frac{1}{2}|E_0|^2\left( {1\pm g^{(1)}(\Delta t) \cos{\omega_0 \Delta t} }\right).
$$
Averaging over $\Delta t$, provided that $g^{(1)}(\Delta t)$ varies negligibly on the scale of fluctuations of $\Delta t$, leads to the expression:
$$
\left<{I_{3}\cdot I_{4}}\right>=\frac{1}{4}|E_0|^4 \left( {1- \left|{g^{(1)}(\Delta t)}\right|^2 \left<{{\cos}^2{\omega_0 \Delta t}}\right> }\right)=\frac{1}{4}|E_0|^4 \left( {1- \frac{1}{2}\left|{g^{(1)}(\Delta t)}\right|^2  }\right)
$$

In full agreement with the Equation ~\ref{eq:15}. The final expression for the correlation function is as follows:


\begin{equation}
    g^{(2)}_{HOM}(\tau,\Delta t)=\frac{1-\frac{1}{2}|g^{(1)}(t_{min})|^2}{1-\frac{1}{2}|g^{(1)}(\Delta t)|^2}
\end{equation}

\subsection{Elliptically polarized pump}

In the case of elliptically polarized excitation with a small ellipticity, the condensate PL also turns out to be elliptically polarized. Moreover, the $\sigma_+$ and $\sigma_-$ PL modes, in addition to intensity imbalance, have frequency detuning, which leads to the beating within the correlation function~\cite{PhysRevLett.128.087402}. The experiment was carried out for pumping powers significantly above the condensation threshold, thus the emission of both polarization modes obtains Poissonian statistics. In order to model this case, we describe the PL in the following form:

\begin{equation} \label{eq:17}
 \begin{pmatrix} 
E_H(t) \\ E_V(t)
\end{pmatrix} =|E_0|e^{ i\varphi(t)+i\varphi_0(t)}\Big[
\begin{pmatrix} 
1 \\ i
\end{pmatrix} 
e^{-i\omega_+ t} + r
\begin{pmatrix} 
1 \\ -i
\end{pmatrix} 
e^{-i\omega_- t}\Big],
\end{equation}

Here $\omega_+$ and $\omega_-$ are natural frequencies of $\psi_+$ and $\psi_-$ polarization modes of PL and $\Delta\omega=\omega_+-\omega_-$ is the frequency of Larmor precession.The $r$ factor reflects the mode imbalance $(0\leq r \leq 1)$. The phase noise $\varphi(t)$ describes the width of both lines ($\omega_{+,-}$), while the decoherence of spin precession will be taken into account when averaging the $\Delta\omega$ value over its width. 
Moreover, for such pumping powers, additional slow signal modulation is observed in the correlation functions for HOM measurements, and therefore we introduced an additional deterministic phase term $\varphi_0(t)=\varphi_0\cos\Omega t$, which is associated with slow adiabatic modulation of the natural frequency $\omega(t)=\omega_0+\varphi_0 \Omega \sin \Omega t$, where $\Omega\ll \omega_0$ and $\varphi_0 \sim 1$. The physical source of this modulation may be related to coherent oscillations of compression and stretching in the sample, to which the excitonic part of the polariton is particularly sensitive via deformation potential \cite{doi:10.1126/science.adn7087}, or the cause may lie somewhere in the experimental setup. We introduce this term to better align the experimental data with the theory, leaving a detailed study of its origin beyond the scope of this article. It is worth noting that in all subsequent expressions, $\varphi_0(t)$ term noticeable affects the shape of the correlation function only for $|\tau|\geq 2~\text{ns}$. Thus, all conclusions regarding the dynamics of the HOM dip in the vicinity of $\tau=0$ for arbitrary $\Delta t$ remain valid regardless of the introduced term.  

Turning to the horizontal component, we obtain the following expression for emission:

\begin{equation}
    E(t)=|E_0|e^{-i\omega_+ t + i\varphi(t)+i\varphi_0(t)}\left({1 + r e^{-i\Delta \omega t }}\right)
\end{equation}

To describe the HOM effect, two functions must be calculated again. For the HBT scheme we have:

\begin{equation}
    G^{(2)}(\tau)=\Big<E^\dagger(t)E^\dagger(t+\tau)E(t+\tau)E(t)\Big>=
\end{equation}
$$
=|E_0|^4\Big<{(1 + r e^{i\Delta \omega t })(1 + r e^{i\Delta \omega (t+\tau) })(1 + r e^{-i\Delta \omega (t+\tau) })(1 + r e^{-i\Delta \omega t })}\Big>=
$$
$$
=|E_0|^4\Big(1+2r^2(1+\cos{\Delta \omega \tau})+r^4 \Big)
$$

For a multi-time term, we get:

\begin{equation}
    G^{(2)}(0,\tau+\Delta t,\tau, \Delta t) =\Big<E^*(t)E^*(t+\tau+\Delta t)E(t+\tau)E(t+\Delta t)\Big>=
\end{equation}
$$
=|E_0|^4\Big<e^{i\varphi(t)+i\varphi(t+\tau+\Delta t)-i\varphi(t+\tau)-i\varphi(t+\Delta t)}\Big>\times\Big<e^{i\varphi_0(t)+i\varphi_0(t+\tau+\Delta t)-i\varphi_0(t+\tau)-i\varphi_0(t+\Delta t)}\Big>
$$
$$
\times\Big<{(1 + r e^{i\Delta \omega t })(1 + r e^{i\Delta \omega (t+\tau+\Delta t) })(1 + r e^{-i\Delta \omega (t+\tau) })(1 + r e^{-i\Delta \omega (t+\Delta t) })}\Big>=
$$
$$
=|E_0|^4\left|{g^{(1)}(t_{min})}\right|^2\times K(\tau,\Delta t) \times\left({
1+2r^2(\cos{\Delta \omega \Delta t} + \cos{\Delta \omega \tau})+r^4
}\right)
$$

where $K(\tau,\Delta t)=J_0\left({4\phi_0 \sin(\frac{1}{2}\Omega\tau)\sin(\frac{1}{2}\Omega\Delta t)}\right)$ is the factor resulting from frequency modulation, that appears together with the square of the coherence function, effectively changing its shape. 

Substituting these results in ~\ref{eq:6} and performing algebraic transformations, we obtain an expression for the correlation function in the HOM scheme:


 \begin{equation}
    G^{(2)}_{HOM}(\tau,\Delta t)=\frac{1}{4}|E_0|^4 \Big[ (1+r^2)^2(1-\frac{1}{2}|g^{(1)}(t_{min})|^2 K(\tau,\Delta t))+
\end{equation}
$$
+2r^2 \sin^2 (\frac{\Delta \omega \Delta t}{2}) |g^{(1)}(t_{min})|^2K(\tau,\Delta t)+
$$
$$
+2r^2 \cos(\Delta \omega \tau) \left({\cos^2 (\frac{\Delta \omega\Delta t}{2}) -\frac{1}{2}|g^{(1)}(t_{min})|^2K(\tau,\Delta t) }\right)\Big]
$$

\subsection{Linearly polarized pump}

In the case of linear polarization of the excitation, the condensate PL experiences a competition of two linear polarization modes. Due to the local anisotropy of the refractive index, $XY$-splitting occurs in the sample, and one of the modes, say the vertical $E_V$, turns out to be preferable and has Poisson statistics, with properties and correlation functions similar to those considered in the case of circular polarization of excitation. And the second horizontal mode $E_H$ is highly suppressed, and act as a source of bunched photons. This suppressed mode is filtered out by the polarizer and measured during the experiment. To account for both intensity and phase oscillations, we describe the condensate emission field as $E(t)=\mathcal{E}(t)e^{i\omega_0 t}$, where the complex amplitude $\mathcal{E}(t)$ obeys Gaussian statistics. Thus, for a complete description of the system, we only need to fix the type of the coherence correlation function of the first order:

\begin{equation}
    g^{(1)}(\tau)=\frac{1}{|E_0|^2} G^{(1)}(\tau)= \frac{1}{|E_0|^2}\left<{\mathcal{E}^\dagger(t+\tau)\mathcal{E}(t)}\right>=\gamma e^{-|\tau|/\tau_0}
\end{equation}

where $|E_0|^2$ is the PL intensity, $\tau_0$ is the coherence time, and we have again introduced the $\gamma$ factor  to account for the incoherent part. And all other second-order correlators for Gaussian statistics are expressed in terms of this function, using the well-known ratio \cite{klauder2006fundamentals,MICHALOWICZ20111233}:

\begin{equation}
    G^{(2)}(t_1,t_2,t_3,t_4)=G^{(1)}(t_1-t_3)G^{(1)}(t_2-t_4)+G^{(1)}(t_1-t_4)G^{(1)}(t_2-t_3)    
\end{equation}

For example, for the correlator in the HBT scheme measurement, the expression is obtained:

\begin{equation}
     G^{(2)}(\tau)=G^{(2)}(t,t+\tau,t+\tau,t)=   
\end{equation}
$$
=\left|{G^{(1)}(\tau)}\right|^2+\left|{G^{(1)}(0)}\right|^2=|E_0|^4\Big(1+\gamma^2 e^{-2|\tau|/\tau_0} \Big)
$$

And for the multi-time term in ~\ref{eq:6}, we get:

\begin{equation}
    G^{(2)}(0,\tau+\Delta t,\tau, \Delta t) =\left|{G^{(1)}(\tau)}\right|^2+\left|{G^{(1)}(\Delta t)}\right|^2=
\end{equation}
$$
=|E_0|^4\Big(\gamma^2 e^{-2|\Delta t|/\tau_0}+\gamma^2 e^{-2|\tau|/\tau_0} \Big)
$$

Substituting these results in ~\ref{eq:6} and by reducing similar terms, we finally obtain the expression for correlation function in HOM scheme:

\begin{equation}
    G^{(2)}_{HOM}(\tau,\Delta t)=\frac{1}{4}|E_0|^2\Big( 1-\frac{1}{2} |g^{(1)}(\Delta t)|^2+\frac{1}{4}|g^{(1)}(\Delta t+\tau)|^2 + \frac{1}{4}|g^{(1)}(\Delta t-\tau)|^2 \Big)
\end{equation}

The normalization in this case is calculated in full accordance with the case of circularly polarized excitation. Thus, the normalized correlation function reads as follows:

\begin{equation}
    g^{(2)}_{HOM}(\tau,\Delta t)= 1+\frac{1}{4}\frac{|g^{(1)}(\Delta t+\tau)|^2 + |g^{(1)}(\Delta t-\tau)|^2}{1-\frac{1}{2} |g^{(1)}(\Delta t)|^2}
\end{equation}

Additionally, in this case, we calculated a correlation function for the case of the absence of optical delay fluctuations $\Delta t$. To do this, it is necessary to keep all the terms in equation $(4)$, expand them according to $(23)$, and after reduction and normalization, the following expression is obtained:

\begin{equation}
    g^{(2)}_{HOM}(\tau,\Delta t, \Phi)= 1+\frac{1}{4}\frac{|g^{(1)}(\Delta t+\tau)|^2 - 2\cos (2\Phi)
g^{(1)}(\Delta t-\tau) g^{(1)}(\Delta t+ \tau)+ |g^{(1)}(\Delta t-\tau)|^2}{1-\frac{1}{2}(1+\cos(2\Phi)) |g^{(1)}(\Delta t)|^2}
\end{equation}

\section{Degradation of Hong-Ou-Mandel effect due to the spin noise}

\begin{figure}[t!]
\includegraphics [width=0.6\columnwidth]{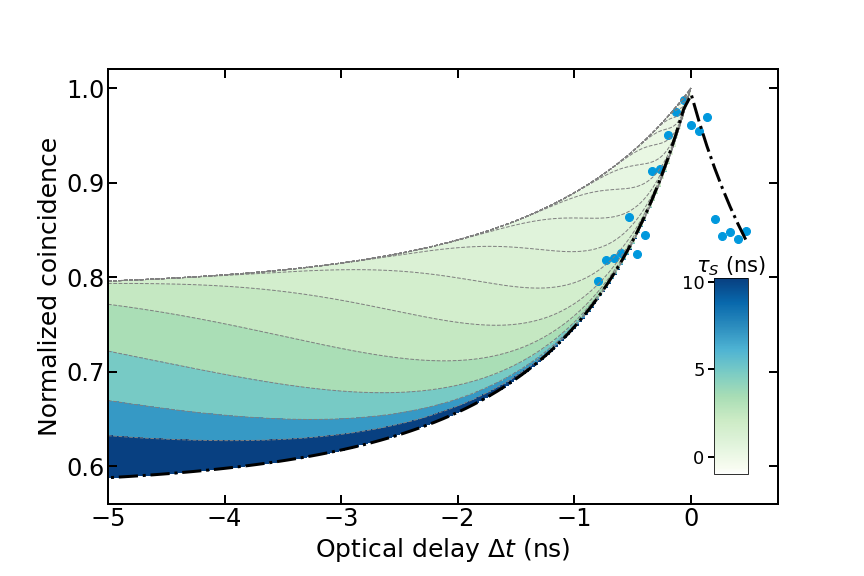}
\centering
\caption{Measured depth of HOM dip at $\tau = 0$ (blue circles) and corresponding theoretical prediction (dash dotted line) with the respect to optical delay $\Delta t$, same as shown in Fig.1(d) of the main text. The gradient depicts how the HOM effect diminishes with spin noise.}
\label{SMfig.4}
\end{figure} 

The visibility of the Hong-Ou-Mandel effect is prominent for being polarization sensitive and is subjected to degradation when the polarizations at the input ports of the beam splitter are mismatched. In order to envision the degradation of the HOM effect due to the spin instability we consider a simple model of the condensate spin noise. We note here that the total intensity of the condensate is constant. We describe the photoluminescence of the condensate similiraly to ~\ref{eq:9}:

\begin{equation}
    E(t)=\boldsymbol{\varepsilon}(t)\cdot |E_0|e^{-i\omega_0 t + i\varphi(t)},
\end{equation}

where $\boldsymbol{\varepsilon}(t)$ is Jones vector parameterizing the polarization as follows:
$$
    \boldsymbol{\varepsilon}(t)=
    \begin{pmatrix} 
    \cos\theta(t) \\ \sin\theta(t)e^{i\phi(t)}
    \end{pmatrix}. 
$$

Parameters $\theta(t)$ and $\phi(t)$ are stochastic variables, describing polarization fluctuations. We assume that this fluctuations are independent of phase fluctuations 
$\varphi(t)$, thus the correlation functions are factorized as follows:

\begin{equation}
    G^{(2)}(t_1,t_2,t_3, t_4) =\Big<E^\dagger(t_1)E^\dagger(t_2)E(t_3)E(t_4)\Big>=
\end{equation}
$$
    =\Big<\big(\boldsymbol{\varepsilon}^\dagger(t_1)\boldsymbol{\varepsilon}(t_4)\big)\big(\boldsymbol{\varepsilon}^\dagger(t_2) \boldsymbol{\varepsilon}(t_3)\big) \Big>\times |E_0|^4\Big<e^{i\varphi(t_1)+i\varphi(t_2)-i\varphi(t_3)-i\varphi(t_4)}\Big>
$$

This equation can be used to calculate the equation ~\ref{eq:6}, where last factor is already obtained in previous section 2A. 
At any given time $\boldsymbol{\varepsilon}^\dagger(t)\cdot \boldsymbol{\varepsilon}(t)=1$. Therefore, the HBT correlation function $G^{(2)}(\tau)$ remains unchanged. Further for simplicity, we will only consider the HOM dip change at $\tau=0$, at various optical delays $\Delta t$. As such we only need to calculate one term:

\begin{equation}
    G^{(2)}(0,\Delta t,0, \Delta t) =\Big<|\boldsymbol{\varepsilon}^\dagger(t)\boldsymbol{\varepsilon}(t+\Delta t)\big|^2 \Big>|g^{(1)}(\Delta t)|^2
\end{equation}

Inspired by the case where $\theta(t)$ has a Gaussian distribution and calculations can be done explicitly, we establish the following qualitative expression for the polarization correlation:

\begin{equation}\label{eq:32}
\Big<|\boldsymbol{\varepsilon}^\dagger(t)\boldsymbol{\varepsilon}(t+\Delta t)\big|^2 \Big>=\frac{1}{2}+
\frac{1}{2}e^{-({\Delta t / \tau_S})^2}
\end{equation}

where  $\tau_{s}$ represents the time for condensate polarization to fluctuate to a random orientation. For $\Delta t> \tau_{s}$ the correlation function \ref{eq:32} approaches $\frac{1}{2}$, which corresponds to the interference of fully polarized light with unpolarized. Thus the equation ~\ref{eq:7} is transformed to

\begin{equation}
    g^{(2)}_{HOM}(\tau=0,\Delta t)=\frac{1-\Big(\frac{1}{4}+
\frac{1}{4}e^{-({\Delta t / \tau_S})^2} \Big)|g^{(1)}(0)|^2}{1-\Big(\frac{1}{4}+
\frac{1}{4}e^{-({\Delta t / \tau_S})^2} \Big)|g^{(1)}(\Delta t)|^2}
\end{equation}

The change of the HOM visibility is shown in Fig.~\ref{SMfig.4} as a gradient  Such an interpretation qualitatively describes the possible discrepancy of the HOM dip from the measured value caused by polarization instabilities.

\section{Hong-Ou-Mandel effect for the condensate confined in traps of different sizes}

\begin{figure}[t!]
\includegraphics [width=0.8\columnwidth]{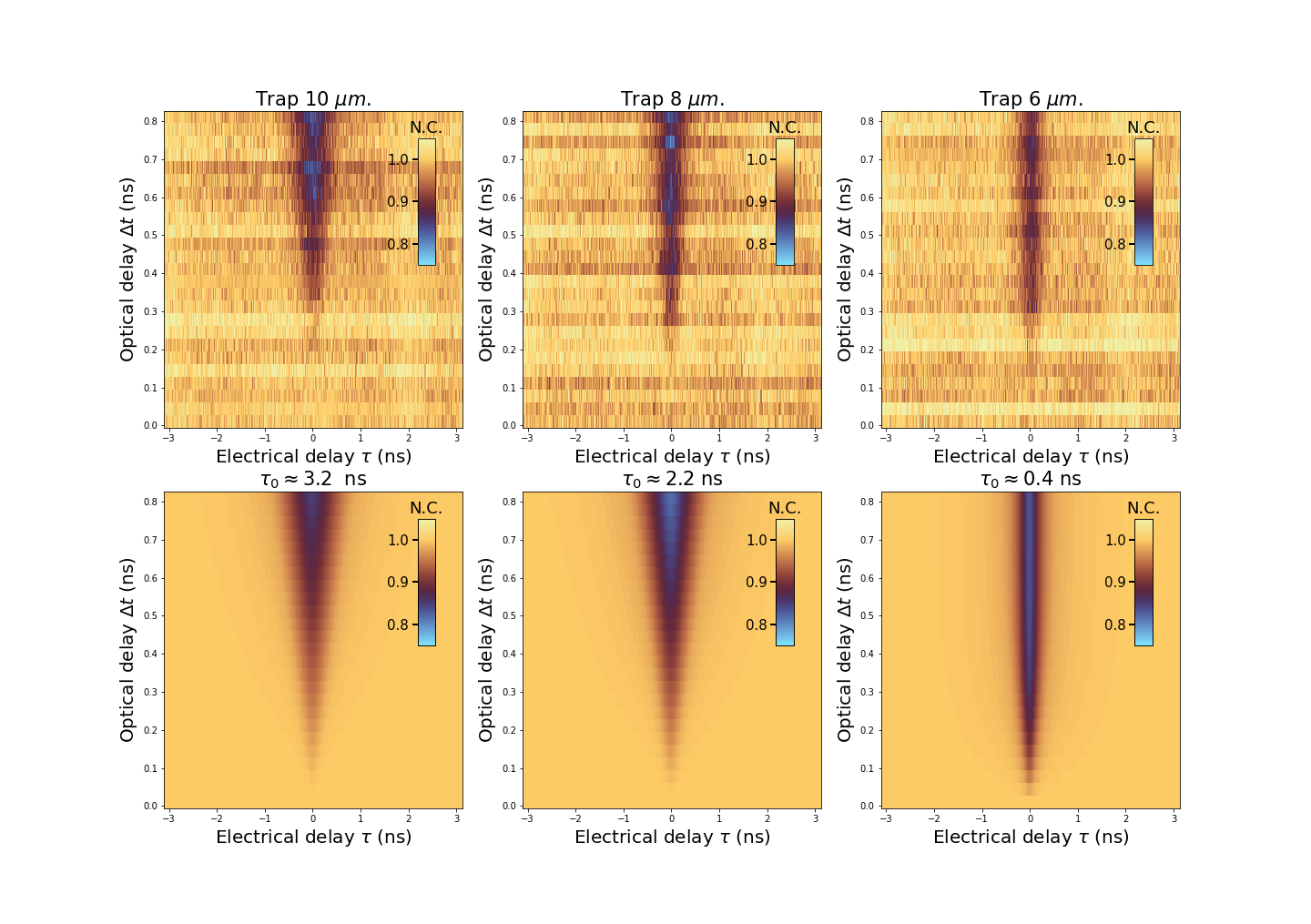}
\centering
\caption{HOM effect color map for the condensate in the trap of three different sizes. The top line is the measured correlation functions, the bottom line is the correlation functions calculated from the theoretical model with fitted coherence times. }
\label{SMfig.3}
\end{figure}

For circularly polarized excitation, the landscape of the HOM effect color map for the condensate in the trap is determined by the photoluminescence first-order coherence function, which is connected to the size of the trap. Smaller the trap size, the more significantly the condensate wave function overlaps with the exciton reservoir, which leads to an effective increase in noise level and a decrease in coherence time. In figure~\ref{SMfig.3} we show a normalized correlation functions measured in HOM interferometer for  traps of three different sizes as well as numerically calculated. Theoretical color maps with adjusted coherence times are in good agreement with experimental data and confirm the stated dependence. We note here that the condensate energy was always at the trap ground state.

\section{Larmor precessions of the trapped polariton condensate pseudo-spin}

\begin{figure}[h!]
\includegraphics [width=0.5\columnwidth]{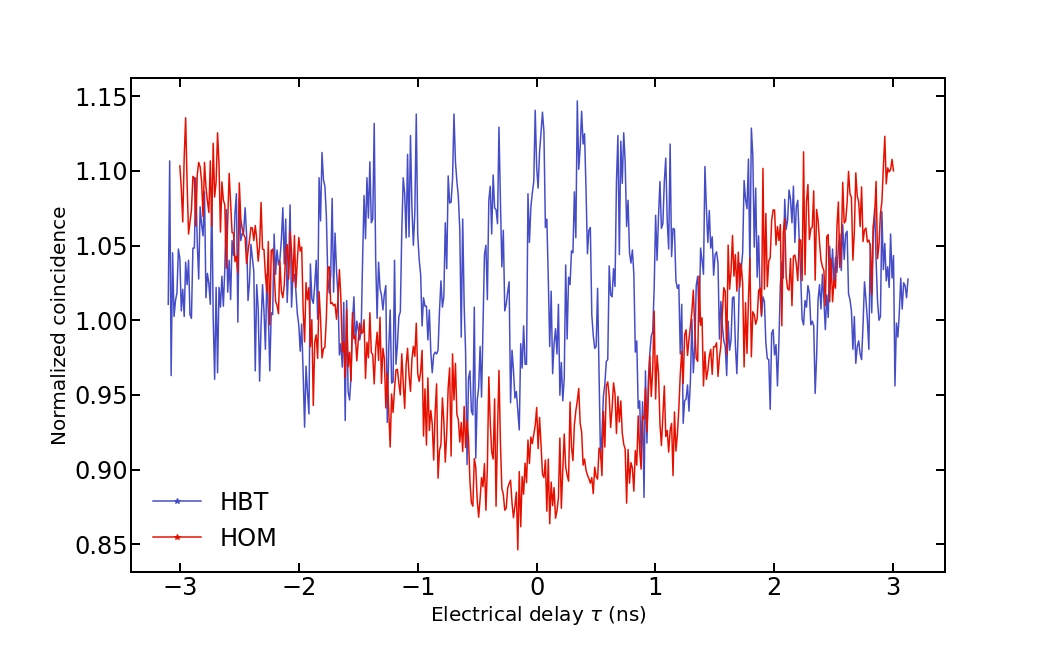}
\centering
\caption{Larmor fast oscillations manifested in second-order correlation functions measured in the HBT scheme (blue) and in the HOM scheme (red) for optical delay $\Delta t=680$ps.}
\label{SMfig.2}
\end{figure}

The periodic high frequency intensity modulation in polarization filtered condensate emission due to spinor Larmor precession is a main factor causing the observed revival of the HOM effect. In figure~\ref{SMfig.2} we show a normalized correlation function measured in HBT interferometer. This data corresponds to data shown in Fig. 2(c) in the main text and underline the high frequency matching between two measurements (HBT and HOM).

\end{document}